\documentclass[11pt]{article}

\usepackage{amssymb}
\usepackage{setspace}
\usepackage{color}
\usepackage{graphicx}

\def\urltilda{\kern -.15em\lower .7ex\hbox{\~{}}\kern .04em}

\title{On the informativeness of dominant and co-dominant genetic markers for Bayesian supervised clustering}

\author{Gilles Guillot\footnote{Department of Informatics and Mathematical Modelling, 
Technical University of Denmark, 
2800, Lyngby, Copenhagen,  Denmark} $\;$ and 
Alexandra Carpentier-Skandalis\footnote{Centre for Ecological and Evolutionary Synthesis, 
Department of Biology, University of Oslo, 
P.O. Box 1066 Blindern, 0316 Oslo, Norway}}

\begin{document}

\maketitle

\baselineskip26pt

\begin{abstract}
We study the accuracy of Bayesian supervised method used to cluster individuals into genetically 
homogeneous groups on the basis of  dominant or codominant molecular markers. 
We  provide a formula relating an error criterion  the number of loci used and the number of 
clusters. 
This formula is exact and holds for arbitrary number of clusters and markers. 
Our work suggests that  dominant markers studies can achieve an accuracy  similar to that 
of   codominant markers studies if  the number of  markers used in the former is about 1.7 times larger 
than in the latter. 
\end{abstract}



\section{Background}\label{sec:background}
A common problem in population genetics consists in assigning an individual to one of $K$ 
populations on the basis of its genotype  and information about the distribution  
of the various alleles in the $K$ populations. 
This question has received a considerable attention in the population genetics and molecular ecology literature 
\cite{Rannala97,Cornuet99,Paetkau04,Piry04} as it can provide important insight about 
gene flow patterns and 
migration rates. It is for example widely used in epidemiology to detect the origin 
of a  pathogens or of their hosts (see e.g. \cite{Gladieux08,Perez08,Bataille09} for examples) 
or in conservation biology and population management to detect illegal 
trans-location or poaching \cite{Manel02}. 
See \cite{Manel05} for a review of related methods.

In a statistical phrasing, assigning an individual to some known clusters 
is a supervised clustering problem. 
This requires  to observe the genotype of the individual to be assigned 
and those of some individuals in the various clusters. 
For diploid organisms 
(i.e. organisms harbouring two copies of each chromosome),  
certain lab techniques allow one to retrieve the exact genotype of each individual. 
In contrast, for  some markers 
it is only possible to say whether a certain allele $A$ (referred hereafter as to dominant allele) 
is present or not at a locus. In this case, one can not distinguish the heterozygous genotype $Aa$ 
from the homozygous genotype $AA$ for the dominant allele. 
The former type of markers are said to be codominant while the latter are said to be dominant.
It is clear that the the second genotyping method incurs a loss of information.
The consequence of this loss of information has been studied from an empirical point of view 
\cite{Campbell03} but it has never been studied on a theoretical basis. 
The choice to use one type of markers  for empirical studies is therefore 
often motivated mostly by practical considerations rather than by an objective rationale 
\cite{Schlotterer04,Bonin07}.
The objective of the present article is to compare the accuracy achieved with dominant and codominant markers 
when they are used to perform supervised clustering and to derive some recommendations 
about the number of markers required to achieve a certain accuracy. 
Dominant markers are essentially bi-allelic in the sense that they record the presence of the absence of a 
certain allele. We are not concerned here by the relation between informativeness
and the level of polymorphisms (cf \cite{Rosenberg01,Kalinowski05} for references on this aspect). 
We therefore  focus on bi-allelic dominant and co-dominant markers.
Hence our study is representative of Amplified Fragment Length Polymorphism (AFLP) and Single Nucleotide Polymorphism (SNP) 
markers, which are some of the most employed markers in genetics.


\section{Informativeness of dominant and co-dominant markers}\label{sec:algo}
\subsection{Cluster model}
We will consider here the case of diploid organisms at $L$ bi-allelic loci. 
We denote by $z=(z_{l})_{l=1,...,L}$ the genotype   of an individual.
We denote by $f_{kl}$ the frequency of allele $A$ in cluster $k$ at locus $l$ 
We assume that each cluster is at Hardy-Weinberg equilibrium (HWE) at each locus.
HWE is defined as the conditions under which the allele carried at a locus on one chromosome 
is independent of the allele carried at the same locus on the homologous chromosome.
This situation  is observed  at neutral loci when individuals mate at random in a cluster. 
Denoting by $z_l$ the number of copies of allele $A$ carried by an individual, we have:
For co-dominant markers, this can be expressed as
\begin{eqnarray}
p(z_l=2|f)  & = & f_l^2 \label{eq:HWEc_begin}\\
p(z_l=1|f) & = & 2f_l(1-f_l)  \\
p(z_l=0|f) & = & (1-f_l)^2\label{eq:HWEc_end}
\end{eqnarray} 
For dominant markers, $z_l$ is equal to 0 or 1 depending on whether a copy of allele $A$ is present 
in the genotype of the individual. Under HWE we have:
\begin{eqnarray}
p(z_l=1|f) & = & f_l^2+2f_l(1-f_l)\label{eq:HWEd_begin}\\
p(z_l=0|f) & = & (1-f_l)^2 \label{eq:HWEd_end}
\end{eqnarray} 
In addition to HWE, we also assume that the various loci are at linkage equilibrium (henceforth HWLE), i.e. 
that the probability 
of a multilocus genotype is equal to the product of probabilities of single-locus genotypes: 
$p(z_1,....,z_L) = \prod_l p(z_l)$. 
We assume that the individual to be classified has origin in one of the $K$ clusters (no admixture).

\subsection{Sampling model}

We will measure the accuracy of a classifying rule for a given type of markers 
by the probability to assign correctly 
an individual with unknown origin. 
We are interested in deriving results that are independent  (i) on the particular origin $c$  of the individual to be classified 
(ii) on the  genotype $z$ of this individual and (iii) 
on the allele frequencies $f$ in the various clusters. 
We will therefore derive results that are conditional on $c$, $z$ and $f$ and then compute Bayesian averages 
under suitable prior distributions.
The  mechanism assumed in the sequel is as follows
\begin{enumerate}
\item The individual has  origin in one of the $K$ clusters. This origin is unknown 
and all origins are equally likely. 
We therefore assume a uniform prior for $c$ on $\{1,..., K\}$.
\item In each cluster, for each locus the allele frequencies 
follow a Dirichlet(1,1) distribution with independent across 
clusters and loci.
\item Conditionally on $c$ and $f$, the probability of the genotype of the individual 
is given by equations (\ref{eq:HWEc_begin}-\ref{eq:HWEc_end}) or (\ref{eq:HWEd_begin}-\ref{eq:HWEd_end}), i.e 
we assume that the individual has been sampled at random among all individuals in his cluster of origin. 

\end{enumerate}

\subsection{Accuracy of assignments under a maximum likelihood principle}

We consider an individual of unknown origin $c$ with known genotype $z$ with potential origin in 
$K$ clusters with known allele frequencies.
Following a maximum likelihood principle, it is natural to estimate $c$ 
as the cluster label for which the probability of observing 
this particular genotype is maximal. 
Formally: $c^* = Argmax_k p(z |  c=k,f_k)$. 
This assignment rule is deterministic, 
but whether the individual is correctly assigned 
will depend on its genotype and on cluster allele frequencies. Randomising these quantities 
and averaging over all possible values, we can derive a generic formula for 
the probability of correct assignment $p^{MLA}$ as 
\begin{eqnarray}
p^{MLA}   & = &    \int_{\varphi}  \sum_{\zeta}   \max_\gamma p(c=\gamma, z=\zeta | f=\varphi)  
d p(\varphi) \label{eq:eqgen}
\end{eqnarray}
See section \ref{sec:MLE_app} in appendix for details. This formula is of little practical use and 
deriving some more explicit expression  for arbitrary value of 
$K$ and $L$ seems to be out of reach.
However, for $K=2$ and $L=1$, under the assumptions that 
the individual has {\em a priori} equally likely ancestry in each cluster 
and that each $f_k$ has a Dirichlet distribution with parameter $(1,...,1)$ (flat).
we get 
\begin{equation}\label{eq:ML_codom}
p^{MLA}_c (K=2,L=1) = 17/24 \mbox{ for  codominant markers}
\end{equation}
 and 
\begin{equation}\label{eq:ML_dom}
p^{MLA}_d (K=2,L=1) = 16/24 \mbox{ for dominant markers. }
\end{equation}

Because of the lack of practical usefulness of eq. (\ref{eq:eqgen}), 
we now define an alternative rule for assignment that is similar in spirit to maximum likelihood but also 
leads to more tractable equations.

\subsection{Accuracy of assignments under a stochastic  rule} 
Considering the collection of likelihood values $p(z |  c=k,f_k)$ for $k=1,...,K$, 
following \cite{Rosenberg03},  we define a stochastic assignment (SA) rule 
by assigning the individual to a group at random with probabilities proportional to 
$p(z |  c=k,f_k)$.
In words, an individual with genotype $z$ 
is randomly assigned to cluster $k$ with a probability proportional 
to the probability to observe this genotype in cluster $k$. 
The rationale behind this rule is that 
high values of   $p(z |  c=k,f_k)$ indicate strong evidence of ancestry in group $k$ but do not 
guarantee against miss-assignments. 
To derive the probability of correct assignment, we first consider that the allele frequencies are known, 
and the account for the uncertainty 
about these frequencies  by Bayesian in integration. 
The use of a Bayesian framework is motivated by the fact that (i) there is genuine uncertainty on allele frequencies 
which can not be overlooked, and (ii) under some fairly mild assumptions,  allele frequencies 
are known to be Dirichlet distributed (possibly with a degree of approximation see e.g. \cite{Guillot08b,Gaggiotti10}). 
Refer to \cite{Beaumont05} for further discussion of the Bayesian paradigm in population genetics.

We now give our main results regarding this clustering rule. 

For bi-allelic loci and denoting by $p^{SA}_c$  
the probability of correct assignment using codominant markers we have:
\begin{equation}\label{eq:prob-codom}
p^{SA}_c (K,L)=  \frac{1}{1+(K-1)(5/8)^L}
\end{equation}

For bi-allelic loci and denoting by $p^{SA}_d$  
the probability of correct assignment using dominant markers is 
\begin{equation}\label{eq:prob-dom}
p^{SA}_d  (K,L)=  \frac{1}{1+(K-1)(25/33)^L} 
\end{equation}

\section{Implications}\label{sec:sup}
Our investigations considered bi-allelic loci and are therefore representative of AFLP and SNP markers 
which are some of the most employed markers in genetics. 
In this context, for supervised clustering, our main conclusions are that 
(i) codominant markers are more accurate than dominant markers, 
(ii) the difference of accuracy decreases toward 0 as 
the number of markers $L$ increases, 
(iii)  $L_d$ dominant markers can achieve an accuracy  even higher than that of  $ L_c$ codominant markers as long as 
the numbers of loci used satisfy $L_d \geq  \lambda  L_c$ where $\lambda = \ln(5/8)/\ln(25/33) \approx 1.69$.

The figures reported have to be taken with a grain of salt as they may depend on some specific aspects of the 
models considered. For example, the model considered here assumes independence of allele frequencies 
across clusters. This assumption is relevant in case of populations displaying low migration rates 
and low amount of shared ancestry. When one of these assumptions is violated, an alternative  parametric model 
based on the Dirichlet distribution that accounts for  correlation of allele frequencies across population is often used 
(see \cite{Guillot08b} and references therein). 
It is expected that the accuracy obtained with both markers would be lower under this model.
Besides, the present study does not account for ascertainment bias 
\cite{Nielsen03,Nielsen04,Foll08,Guillot09c}, an aspect that might affect the results but is notoriously difficult 
to deal with. 
However, it is important to note that the conditions considered in the present study  
were the same for dominant and codominant markers 
so that results should  not be biased toward one type of marker.
Our global result about the relative informativeness of dominant and co-dominant markers 
contrasts with the common belief  that dominant markers are expedient 
one would resort to when co-dominant markers are not available (see \cite{Bonin07} for discussions). \\
A comparison of dominant and codominant markers for unsupervised clustering 
has been carried out  \cite{Guillot10b}. This study based on simulations suggests 
that the loss of accuracy incurred by dominant markers in unsupervised clustering 
is much larger than for supervised clustering. 
This is presumably explained by the fact that in case of HWLE clusters, supervised clustering 
seeks to optimise a criterion based on allele frequencies only. 
This contrasts with unsupervised clustering which seeks to optimise a criterion based on allele frequencies 
and HWLE. 
A  similar theoretical analysis of unsupervised clustering algorithm similar to the present study 
would be valuable but we anticipate that it would present more difficulties.

\section*{Acknowlegement}
This work has been supported by Agence Nationale de la Recherche grant ANR-09-BLAN-0145-01.

\clearpage 
\appendix

\section{Supervised clustering with a maximum likelihood principle}\label{sec:MLE_app}
We consider the setting where the unknown ancestry $c$ of an individual with genotype $z$ is estimated by  
$c^* = \mbox{Argmax}_c p(z | c,f)$. 
As this estimator is a deterministic function of $z$ we denote it by $c^*_z$ for clarity in the sequel.
Consider for now that the allele frequencies $f$ are known to be equal to some $\varphi$.  
Under this setting, randomness comes from the sampling of $c$ and then from the sampling of $z|(c,f)$.
We are concerned with the event ${\cal E}$ defined  as $${\cal E} =\{\mbox{the individual is correctly assigned} \}$$. 
Applying the total probability formula, we can write 
\begin{equation}
p({\cal E}|f=\varphi)  = \sum_{\gamma} \sum_{\zeta} p({\cal E},c=\gamma,z=\zeta|f=\varphi) 
\end{equation}
In the sum over $\gamma$, only one term is not equal to $0$, this is the term for $\gamma=c^*$, hence

\begin{eqnarray}
p({\cal E}|f=\varphi)  & = &  \sum_{\zeta} p({\cal E},c=c^*_{\zeta},z=\zeta|f=\varphi) \\
& = & \sum_{\zeta} p(c=c^*_{\zeta},z=\zeta|f=\varphi)\\
& = & \sum_{\zeta} p(c=c^*_{\zeta}|f=\varphi) p(z=\zeta|c=c^*_{\zeta},f=\varphi)\\
& = & \sum_{\zeta} p(c=c^*_{\zeta}) p(z=\zeta|c=c^*_{\zeta},f=\varphi)
\end{eqnarray}

Assuming that the individual has {\em a priori} equally likely ancestry in each cluster, i.e. 
 assuming a uniform distribution for the class variable $c$, we get 
 \begin{equation}
p({\cal E}|f=\varphi)  =  K^{-1}  \sum_{\zeta}  p(z=\zeta | c=c^*_{\zeta},f=\varphi)
\end{equation}

By definition, $c^*_z$ satisfies $ p(z | c^*_z,f) = \max_\gamma p(z | c=\gamma,f)$, hence
 \begin{eqnarray}
p({\cal E}|f=\varphi)  & = &   K^{-1}   \sum_{\zeta} \max_{\gamma} p(z=\zeta | c=\gamma,f=\varphi) \\
 & = &    \sum_{\zeta}   \max_\gamma p(c=\gamma)  p(z=\zeta | c=\gamma,f=\varphi) \\
 & = &    \sum_{\zeta}   \max_\gamma p(c=\gamma, z=\zeta | f=\varphi)
\end{eqnarray}

We seek an expression of the probability of correct assignment that does not depend on particular 
values of allele frequencies. This can be obtained by integrating over allele frequencies, namely
\begin{eqnarray}
p({\cal E})  & = &   \int_{\varphi} p({\cal E}|f=\varphi) d p(\varphi) \\ 
 & = &    \int_{\varphi}  \sum_{\zeta}   \max_\gamma p(c=\gamma, z=\zeta | f=\varphi)  
d p(\varphi) \label{eq:eqgen_app}
\end{eqnarray}

Note that  identity (\ref{eq:eqgen_app}) holds for any number of cluster $K$, any number of loci $L$ 
and any type of markers 
(dominant vs. codominant).\\

We now consider a two cluster problem in the case where the genotype of an individual has been recorded 
at a single bi-allelic locus. 
We denote by $f_1$ (resp. $f_2$) the frequency of allele $A$ in cluster 1 (resp. cluster 2).

\subsection{Codominant markers:}
There are only three genotypes: $AA,Aa$ and $aa$. 
Denoting by $f_k$ the frequency of allele $A$ in cluster $k$ and 
conditionally on  $f_k$, these three genotypes occur in cluster $k$ with probabilities 
$f_k^2, 2f_k(1-f_k)$ and $(1-f_k)^2$, and equation (\ref{eq:eqgen_app}) can be simplified as
\begin{eqnarray}
p({\cal E})   =  \int_{\varphi}  p(c) 
\left[ \max_{\gamma}  f_\gamma^2 + 2 \max_{\gamma} f_\gamma(1-f_\gamma) + \max_{\gamma} (1-f_\gamma)^2 \right] 
d p(\varphi) 
\end{eqnarray}

We need to derive the distribution of  $\max_{\gamma}  f_\gamma^2$ and of $\max_{\gamma} f_\gamma(1-f_\gamma)$. 
Assuming a flat Dirichlet distribution for $f_k$, elementary computations give:

\begin{equation}
p (\max_k f_k^2 < x) = x
\end{equation}
i.e $\max_k f_k^2$ follows a uniform distribution on $[0,1]$
so that 
\begin{equation}
\mathbb{E} (\max_k f_k^2) = 1/2
\end{equation}

Besides, we also get  
\begin{equation}
p (\max_k f_k(1-f_k) < x) = (1 - \sqrt{1-4x})^2
\end{equation}
and deriving

\begin{equation}
\frac{d p}{dx} (\max_k f_k(1-f_k) < x) = 4 \frac{1 - \sqrt{1-4x}}{\sqrt{1-4x}}
\end{equation}

Integrating by part, we get
\begin{equation}
\mathbb{E} (\max_k f_k(1-f_k)) = \int_0^{1/4} 4x \frac{1 - \sqrt{1-4x}}{\sqrt{1-4x}} dx = 5/24
\end{equation}

Eventually
\begin{eqnarray}
p({\cal E})  & = &  17/24
\end{eqnarray}
which proves equation (\ref{eq:ML_codom})\hfill$\square$.

\subsection{Dominant markers:}

For a single locus, there are two genotypes $A$ and $a$.
Conditionally on  $f_k$, these two genotypes are observed in cluster $k$ with probabilities  
$1-f_k^2$ and $f_k^2$.
Equation (\ref{eq:eqgen_app}) can be simplified here as

\begin{eqnarray}
p({\cal E})   =  \int_{\varphi}  p(c) 
\left[ \max_{\gamma}  f_\gamma^2 + \max_{\gamma} (1-f_\gamma^2) \right] 
d p(\varphi) 
\end{eqnarray}

We now need the density of $1-f_\gamma^2$
\begin{equation}
p( \max_k (1 -f_k^2) < x) = (1-\sqrt{1-x})^2
\end{equation}

\begin{equation}
\frac{d p}{dx} ( \max_k (1 -f_k^2) < x) = \frac{1}{\sqrt{1-x}} - 1
\end{equation}

and 
\begin{equation}
 \mathbb{E}( \max_k (1 -f_k^2) ) = \int_0^1 x\left( \frac{1}{\sqrt{1-x}} - 1 \right) dx = 5/6
\end{equation}

Eventually we get 
\begin{eqnarray}
p({\cal E})  & = &  16/24
\end{eqnarray}
which proves equation (\ref{eq:ML_dom})\hfill$\square$.

\section{Stochastic assignment rule} 
\begin{figure}[h]
\vspace{-.001cm}\hspace{4cm}
\begin{tabular}{lc}
\hspace{-2.2cm}\vspace{0cm}\includegraphics[angle=0,width=3cm]{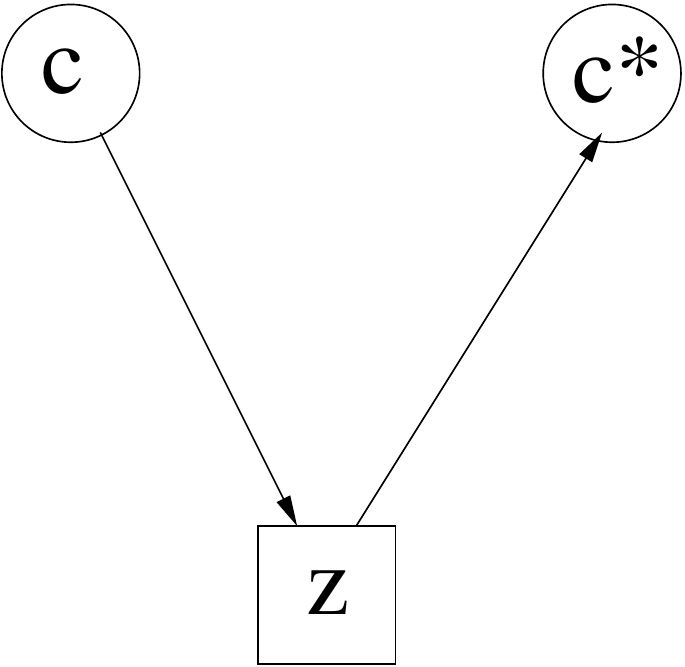}&
\hspace{+2.3cm}\vspace{0cm}\includegraphics[angle=0,width=3cm]{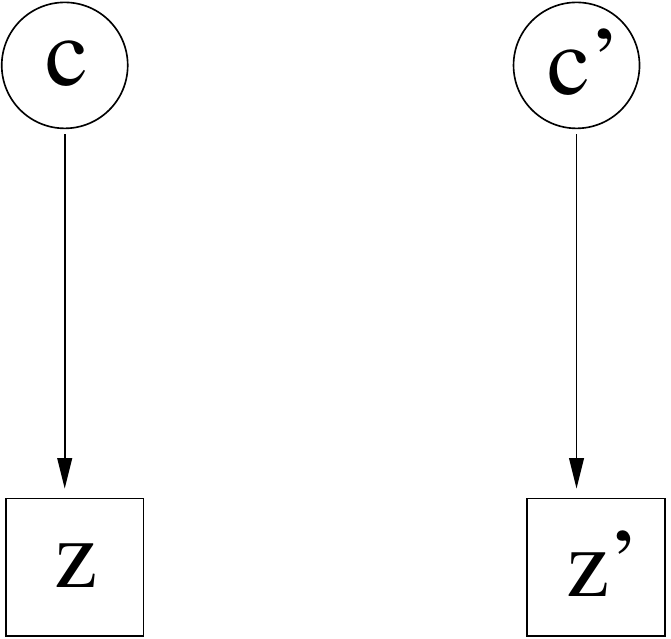}\\
\hspace{-1cm}(a) & \hspace{1.8cm}(b)
\end{tabular}
\vspace{-0cm}
\caption{Directed acyclic graph  for our stochastic assignment rule (left) and for an alternative 
scheme (right).  All downward arrows represent the same conditional dependence given by our likelihood model. 
Upward arrow represents the reverse probability dependence. }\label{fig:dag}
\end{figure}

The maximum likelihood assignment rule considered above is not tractable for arbitrary values of $K$ and 
$L$ (cf. eq. (\ref{eq:eqgen_app})). In particular, a difficulty arises from the maximisation involved. 
We consider here an assignment rule that does not involve maximisation. 
The unknown ancestry $c$ of an individual with genotype $z$ is predicted
by a random variable $c^* $ with values in $\{1,...,K\}$ and such that 
$p(c^*=k | z,f) \propto p( z | c=k,f)$. 
As in the previous sections, 
we first consider that the allele frequencies are known, 
however we skip this dependence in the notation at the beginning for clarity. 
We will account for the uncertainty 
about these frequencies later by Bayesian in integration. 
In this setting, the structure of conditional probability dependence can be represented by 
a directed acyclic graph as in 
the on left-hand side of figure \ref{fig:dag}.

We are concerned with evaluating the probability of  event 
${\cal E}$ defined  as $${\cal E} =\{\mbox{the individual is correctly assigned} \}$$.
i.e. $ {\cal E} = \{c=c^*\}$. 
We denote by $ p_a$ (resp. $ p_b$) probabilities under the two conditional dependence 
structure of figure \ref{fig:dag}.
Some elementary computations show that $p({\cal E})$ can be expressed in terms 
of a probability in the model of the right-hand-side of the DAG in figure \ref{fig:dag}, namely: 
\begin{eqnarray}
p_a(c=c^*)  & = & p_b(c=c' | z=z')   \label{eq:finte_ACS}
\end{eqnarray}
The left-hand-side of this expression can be written as 
\begin{eqnarray}
 p_b(c=c' | z=z') & = & p_b(c=c' , z=z')/ p_b( z=z') 
\end{eqnarray}
It  is more convenient to manipulate this expression than $p_b(c=c^*)$. 
 We will to use it to evaluate $p_a({\cal E})$.

\subsection{Codominant markers:}

We assume  that the individual has {\em a priori} equally likely ancestry in each cluster. 
We slightly change the notation denoting by $z_l$ the count of allele $A$ at locus $l$ for the individual to be assigned. 
Then making the dependence on $f$ explicit in the notation, we have
\begin{eqnarray}
p_b(c=c', z=z'|f )  & = & \sum_z \sum_k p^2_b(c,z| f) \\
  & = & \sum_z \sum_k \left[ \frac{1}{K} \prod_l f_k^{z_l}(1-f_k)^{2-z_l} (2-\delta_{z_l}^1) \right]^2
\end{eqnarray}

where $\delta_{z_l}^1$ denotes the Kronecker symbol that equals 1 if $z_l=1$ and 0 otherwise.

Accounting for uncertainty about $f$ by integration, we get
\begin{eqnarray}
p_b(c=c', z=z')  & = & \int_f p_b(c=c', z=z'|f ) df \\
  & = & \int_f  \sum_z \sum_k \left[ \frac{1}{K} \prod_l f_k^{z_l}(1-f_k)^{2-z_l} (2-\delta_{z_l}^1) \right]^2  df  \label{eq:intf}
\end{eqnarray}

Among the terms enumerated in the sum over $z$ above, let us consider a generic term $z$ for which 
the number of loci having exactly $h$ heterozygous genotypes.
The term corresponding to such a genotype $z$ in the sum above can be written

\begin{equation}
 \sum_k \frac{1}{K^2} 2^{2h}  \left[ \int_f f^2(1-f)^2 df \right]^h 
 \left[ \int_f f^4df \right]^{L-h}
\end{equation}

Denoting by $C_L^h$ the binomial coefficient, there are $\displaystyle C_L^h 2^{L-h}$ such terms. 
Equation (\ref{eq:intf}) becomes
\begin{eqnarray}
p_b(c=c', z=z')  & = & \sum_h \sum_k C_L^h 2^{L-h} \frac{1}{K^2} 2^{2h}  \left[ \int_f f^2(1-f)^2 df \right]^h 
 \left[ \int_f f^4df \right]^{L-h}
\end{eqnarray}

Assuming a flat Dirichlet distribution for the allele frequencies, we get
\begin{eqnarray}
p_b(c=c', z=z')  & = & \frac{1}{K} \left( \frac{8}{15} \right)^L
 \end{eqnarray}

We now need to evaluate $p_b(z=z')$, but since 
\begin{eqnarray}
p_b(z=z' |f)  & = & \sum_z p^2_b(z|f) = \sum_z \left( \sum_k p_b(k,z | f)\right)^2,
 \end{eqnarray}

\begin{eqnarray}
p_b(z=z')  & = & \int_z \sum_z p^2_b(z|f) = \sum_z \left( \sum_k p_b(k,z | f)\right)^2 \\
   & = &   \int_z  \sum_z \left(  \sum_k p_b(k,z | f)^2  + 
\sum_{k\neq k'}  p_b(k,z | f) p_b(k',z | f)\right) \\
  & = & p_b(c=c',z=z' ) + \nonumber \\
  && \sum_h C_L^h 2^{L-h} \left( \sum_{k \neq k'} 2 \left(\frac{1}{K^2}2^{2h} 
\left[ \int_f f(1-f) df \right]^{2h} 
 \left[ \int_f f df \right]^{4(L-h)} \right) \right) \\
  & = & \frac{1}{K} \left( \frac{8}{15} \right)^L + \frac{K-1}{K} \frac{1}{3^L}
 \end{eqnarray}

Eventually,

\begin{eqnarray}
p({\cal E})  & = & \frac{1}{1+(K-1)\left(\frac{5}{8}\right)^L}
 \end{eqnarray}
which proves equation (\ref{eq:prob-codom}). \hfill$\square$

\vspace{1cm}

\subsection{Dominant markers:}
We still have 
\begin{eqnarray}
p_b(c=c', z=z')    & = & 
\int_f  \sum_z \sum_k \left[ \frac{1}{K} \prod_l f_k^{z_l}(1-f_k)^{2-z_l} (2-\delta_{z_l}^1) \right]^2  df  
\end{eqnarray}

For a generic genotype $z$ in the sum above, let us denote  by $r$ the number of loci carrying exactly one copy 
of the recessive allele, then

\begin{eqnarray}
p_b(c=c', z=z')    & = & 
 \sum_r \sum_k C_r^l  \frac{1}{K^2}  \left[ \int_f f_k^4 df \right]^r  
\left[ \int_f (1-f_k^2)^2 df \right]^{L-r}  \\
 & = & \sum_r \sum_k C_r^l \frac{1}{K^2} \left( \frac{1}{5} \right)^L \left( \frac{8}{3} \right)^{L-r} \\
 & = & \frac{1}{K}\left( \frac{11}{15} \right)^L
\end{eqnarray}

Moreover, by arguments similar to those used for codominant markers, we get 
\begin{eqnarray}
p_b(z=z')    & = &   \frac{1}{K}\left( \frac{11}{15} \right)^L
+  \frac{K-1}{K}\left( \frac{5}{9} \right)^L
 \end{eqnarray}

And we get 
\begin{eqnarray}
p_b(z=z')    & = &   \frac{1}{K}\left( \frac{11}{15} \right)^L
+  \frac{K-1}{K}\left( \frac{5}{9} \right)^L
 \end{eqnarray}

Eventually,

\begin{eqnarray}
p({\cal E})  & = & \frac{1}{1+(K-1)\left(\frac{25}{33}\right)^L}
 \end{eqnarray}
which proves equation (\ref{eq:prob-dom}). \hfill$\square$



\end{document}